\documentclass[twocolumn,prl]{revtex4}

\usepackage{amsmath}
\usepackage{amssymb}
\usepackage{latexsym}

\begin{document}

\bibliographystyle{ieeetr}

\title{Biology's next revolution\footnote{This is a slightly expanded version of an invited essay that
appeared in {\sl Nature}, {\bf 445}, 369 (25 Jan 2007).  This version
includes numerous important citations and a grant acknowledgment omitted from the published version due to space
limitations.}}

\author{Nigel Goldenfeld$^1$ and Carl Woese$^2$}

\affiliation{$^1$Department of Physics and Institute for Genomic Biology,
University of Illinois at Urbana-Champaign,
1110 W. Green St., Urbana, IL 61801, USA}

\affiliation{$^2$Department of Microbiology and Institute for Genomic Biology,
601 South Goodwin Avenue, Urbana, IL 61801, USA
}

\begin{abstract}
The interpretation of recent environmental genomics data exposes the
far-reaching influence of horizontal gene transfer, and is changing our
basic concepts of organism, species and evolution itself.
\end{abstract}

\maketitle

One of the most fundamental patterns of scientific discovery is the
revolution in thought that accompanies the acquisition of an entirely
new body of data.  The new window on the Universe opened up by
satellite-based astronomy has in the last decade overthrown our most
cherished notions of Cosmology, especially related to the size,
dynamics and composition of the Universe. Similarly, the convergence of
new theoretical ideas in evolution together with the coming avalanche
of environmental genomic data, especially from marine microbes and
viruses, will fundamentally alter our understanding of the global
biosphere, and is likely to cause a revision of such basic and
widely-held notions as species, organism and evolution.  Here's why we
foresee such a dramatic transformation on the horizon, and how
biologists will need to join forces with quantitative scientists, such
as physicists, to create a biology that embraces collective phenomena
and supersedes the molecular reductionism of the twentieth century.

The place to start is the notion of horizontal gene transfer
(HGT)\cite{ochman2000lgt}, sometimes touted as the \lq\lq spoiler" in
molecular phylogeny\cite{woese2000ats} because of the deviations it
implies from a perfectly structured picture of Darwinian, vertical,
evolution.  At the microbial level, HGT is a pervasive and incredibly
strong force\cite{ochman2000lgt}, famously accelerating the spread of
antibiotic resistance for example\cite{davies1996oae, salyers1997arg}.
The widespread occurrence of HGT means that it is not a good
approximation (except perhaps within the sterile confines of a
laboratory) to regard microbes as organisms that are dominated by
individual characteristics\cite{thomas2005mab}.  In fact, their
communications at both the genetic and signal response levels (think
quorum sensing) indicate that microbial behavior must be understood as
predominantly cooperative.  In the wild---and recall that perhaps only
about one percent of microbes can currently be cultured---microbes form
communities, invade biochemical niches, and partake in biogeochemical
cycles on a global scale.  The available studies strongly indicate that
microbes absorb and discard genes as needed\cite{szpirer1999rog}, and
in response to their environment\cite{lorenz1994bgt,
sorensen2005sph, frigaard2006plg, coleman2006gia}. Rather than a
discrete spectrum, we see a continuum of genomic possibilities, casting
doubt on the validity of the fundamental concept of species, extended
into the microbial realm\cite{whitaker2005rsn}. The lack of usefulness
of the species concept\cite{sonea1988bwl} is inherent in the recent
exciting forays into metagenomics\cite{allen2005cgm, sogin2006mdd}; for
example, studies of the spatial distribution of rhodopsin genes
observed in marine microbes are in accord with the concept of \lq\lq
cosmopolitan genes", wandering among bacteria, or archaea, as
environmental pressures dictate\cite{frigaard2006plg}.

Equally exciting is the growing realization that the virosphere plays
an absolutely fundamental role in the biosphere on both immediate and
long-term evolutionary senses\cite{brussow2004pae, suttle2005vs}.
Recent work, including our own unpublished work, suggests that viruses
may play an important role as a repository and memory of a community's
genetic information, contributing to the evolutionary
dynamics\cite{filee2003rpv, weinbauer2004vdm} and the stability of the
system.  This is hinted at, for example, by environmental triggering of
prophage induction\cite{williamson2002svl}, in which viruses latent in
cells can become activated by environmental influences. The ensuing
destruction of the cell and viral replication is a potent mechanism for
dispersal of host and viral genes.

What is becoming clear is that microorganisms have a
remarkable ability to reconstruct their genomes in the face of dire
environmental stresses\cite{Radman2006}, and that in some cases at
least\cite{furuta1997cvp}, their collective interactions with the
virosphere (and perhaps other gene transfer agents) may be crucial to
this.  In such a situation, how valid is the very concept of an
organism in isolation? It seems that there is a continuity of energy
flux, communication, informational transfer from the genome up through
cells, community, virosphere, and environment.  If the interactions
are strong, and collective effects dominant, then an organism cannot
even be considered in isolation. Indeed, we would go so far as to
suggest that a defining characteristic of life is the strong
dependency on flux from the environment, be it energy-giving,
chemical-giving, metabolism-giving, or genetically-giving.  This
inherently biocomplex perspective renders academic such debates as
\lq\lq is a virus dead or alive?".

Nowhere are the implications of collective phenomena, mediated by HGT,
so pervasive and important as in evolution itself. A computer scientist
might term the cell's translational apparatus (used to convert genetic
information to proteins) an 'operating system', by which all innovation
is communicated and realized. The fundamental role played by
translation, represented in particular by the genetic code, is shown by
the clearly-documented optimization of the code\cite{haig1991qme}. Its
special role in any form of life leads inexorably to the rather
striking prediction that early life must have evolved in an inherently
Lamarckian way, with vertical descent marginalised by the more powerful
early forms of HGT\cite{vetsigian2006cea}. Such gradual refinement
through the horizontal sharing of genetic innovations would have led to
the generation of a combinatorial explosion of genetic novelty, until
the level of complexity, as exemplified perhaps by the multiple levels
of regulation, required a transition to the present era of vertical
evolution.  Thus, we regard as rather regrettable the conventional
concatenation of Darwin's name with evolution, because there are other
modalities that must be entertained and which we regard as mandatory
during the course of evolutionary time.

This is an extraordinary time for biology, because the perspective we
have indicated places biology within a context that must necessarily
engage other disciplines (such as statistical physics) that are more
strongly aware of the importance of collective phenomena than biology
has been until now.

Questions suggested by the generic energy, information and gene flows
to which we have alluded will probably require resolution in the spirit
of statistical mechanics and dynamical systems theory. In time, the
current approach of post-hoc modelling will be replaced by interplay
between quantitative prediction and experimental test, nowadays more
characteristic of the physical sciences.

Sometimes, language expresses ignorance rather than knowledge, as in
the case of the nomenclature \lq\lq prokaryote" now superseded by the
terms Archaea and Bacteria, deriving from detailed and well-founded
molecular phylogeny.  We foresee that in biology, new concepts will
require a new language, one that is grounded in the discoveries
emerging from the new data we have highlighted in this essay.  At the
time of an earlier revolution, Lavoisier\cite{lavoisier2tec} observed
that scientific progress, just like evolution itself, must overcome a
challenge of communication:

\lq\lq ...we cannot improve the language of any science without at the same
time improving the science itself; neither can we, on the other hand,
improve a science, without improving the language or nomenclature which
belongs to it."

Biology is about to meet this challenge.

\noindent
{\bf Acknowledgment} This work is based upon work supported by the
National Science Foundation under grant no. NSF-EF-0526747.

\bibliography{Evolution}
\end{document}